\documentclass[12pt,preprint]{aastex}

\slugcomment{ApJL in press}

\shorttitle{The inclinations of faint TNOs}
\shortauthors{Trilling et al.}

\begin{document}


\title{The inclinations of faint TNOs}


\author{D. E. Trilling\altaffilmark{1}}

\author{C. I. Fuentes\altaffilmark{2,1}}

\and

\author{M. J. Holman\altaffilmark{2}}


\altaffiltext{1}{Department of Physics and Astronomy,
Northern Arizona University, PO Box 6010,
Flagstaff, AZ 86011;
\email{david.trilling@nau.edu}}
\altaffiltext{2}{Harvard-Smithsonian Center for Astrophysics,
    60 Garden Street, Cambridge, MA 02138}


\begin{abstract}
\citet{b04} found that the population of faint
(R$>$26) trans-Neptunian objects (TNOs)
known at that time 
was dominated by ``Classical'' objects,
which have
low inclinations ($i<5^\circ$)
and distances 40--45~AU.
Since those observations, 
the number of faint TNOs whose orbits
are sufficiently well known to be classified
as ``Classical'' or ``Excited''
has grown from seven to~39.
We analyze the dynamical classifications
of faint TNOs known today and find
that 
this population is
dominated by Excited objects.
We discuss some implications of this
result.
\end{abstract}





\section{Introduction}

The trans-Neptunian region is inhabited by
rocky-icy bodies called
trans-Neptunian objects (TNOs). Most TNOs
are faint, and extensive
sky coverage and sensitive surveys
are needed
for their detection.
Understanding the properties of TNOs ---
numbers, sizes,
compositions, and orbital properties ---
helps constrain the history
of the outer Solar System.

The deepest TNO survey performed to date
is that of \citet[][hereafter B04]{b04}, who surveyed
0.02~deg$^2$ to an
F606W magnitude of~29.2 (50\% completion
rate) with the Hubble Space Telescope (HST).
B04 found three new faint TNOs (R~magnitudes
26.3--27.8). Significantly, all three 
of these new faint TNOs have orbits
that lie very close to the ecliptic plane,
all with inclinations less than 2.5~degrees.
B04 compiled all existing TNO survey
data at the time and, in combination with
their three new faint TNOs, concluded that 
faint TNOs were far more likely to be 
low-inclination objects, which they named
``Classical'' TNOs, and far less likely to
be high-inclination (``Excited'') TNOs.
At the time of the B04 survey, there
were only four other R$>$26 TNOs known.

In the years since the B04 survey, two groups
have significantly extended the number of
faint TNOs known, primarily using ground-based
facilities (Table~\ref{datatable}).
There are now 39~known TNOs with R$>$26 and
known inclinations.
It is therefore
possible now to repeat the analysis that
B04 did to determine the relative
prevelance of Classical and Excited TNOs,
at a far higher significance.
We present in this paper our analysis of 
the dynamical state of faint TNOs, using
data available as of June, 2010.

\section{Data and analysis}


We compile data from the surveys listed in
Table~\ref{datatable}.
These are the published surveys that have some sensitivity
to objects with R$\gtrsim25.5$. We additionally
include \citet{elliot05} to provide a ``baseline''
for brighter objects, but our compilation for
objects brighter than R=26 is not meant to 
be complete, just representative.
Most of these papers
report either R magnitudes, or transformations
from their bandpasses to R, so the compiled
data is all in R magnitude as a common
system.
We use the B04 definitions of Classical 
versus Excited, assigning to the former class
all objects with inclinations less than 5~degrees
and to the latter
group all objects with inclinations larger
than 5~degrees. We do not use distance as
a discriminant in the present study; that
parameter is significantly more difficult
to measure in ground-based surveys than it
is with HST.

Figure~\ref{ratio} shows the ratio 
of Classical to Excited objects as a function
of R~magnitude (dark and light
data points). An equal number of Classical
and Excited TNOs would have a ratio of unity.
Of the 648~TNOs in our compendium,
333~are Classical, with the remaining
315~objects belonging to the Excited class.
While the Classical objects slightly dominate overall,
we find that Excited objects dominate in several
magnitude ranges (Figure~\ref{ratio}).
Significantly, Excited objects dominate the
brightest (R$<$22) and faintest
(R$>$25) magnitude ranges, while Classical objects
dominate the middle magnitudes.
The largest deviations from unity
are very large, about 50\%.

We also show in Figure~\ref{ratio} two
best-fit curves from B04: a double-power
law (thick curving line) and a rolling-index
law (thin curving line). These two curves in B04 fit the data
known at that time approximately equally well.
The rolling-index law fits
the present data R$<$24 well,
but the agreement is much less
good at R$>$24, where we have
included significant new data in this work.
The double-power law fits the present
data reasonably well, with the biggest
deviations faintward of R=26, where 
the most significant new data is found.
(We do not include all of the bright-end
surveys that B04 did, so the bright-end agreement
between the B04 curves and the data we
present here is probably satisfactory
due to our bright-end incompleteness.)

Both B04 curves predict, based on fitting the
data known at the time, that the
R$>$26 TNO population should be
dominated by Classical objects.
Clearly, this is because the data fit by B04
includes only seven objects with R$>$26,
three of which are the Classical objects discovered
by B04.
Compared to B04, we include here more than five times
as many objects with R$>$26.
This new compilation shows
that Excited objects dominate this 
faint population.

\section{Possible biases}


With such difficult observations from a range of facilities,
biases could easily be significant.
For example,
most of the new TNOs fainter than R=26 were
discovered in ground-based searches, as compared 
to B04's space-based search. The most obvious
difference between those two techniques is that
ground-based searches have much poorer angular resolution,
and therefore, in general, produce less precise
orbital solutions. If some aspect of less
precise solutions could preferentially produce
Excited orbits then the result presented here
could be explained, perhaps.
As an example, the errors on the inclinations reported
in \citet{fraser08} are typically 10--20~degrees,
which means that the confidence in any individual
classification of Classical may be low, while the classification
of an object as Excited may be higher. This has
the effect of overestimating the number of
Classical TNOs, and therefore does not
rectify the apparent disagreement between
B04 and the present work.
Additionally, 
as
a counter-argument, the reason that B04 used
inclination as the discriminant is because
this orbital element is most easily and
accurately measured in ground-based surveys,
as inclination is manifest simply by the
angle of TNO motion with respect to the 
ecliptic plane. Systematically overestimating
this angle to the degree required to explain
the results presented here is quite unlikely.

Many TNO surveys use multi-lunation follow-up
observations to secure orbits, and these 
long arcs may be subject to biases
that are dependent on dynamical class
\citep{jones2010}.
However, all of the faint TNOs used here
were discovered in pencil beam-type surveys,
which have no long-term recovery observations.
There is essentially no
bias against Classical orbits in such an 
observational scheme.

All of the nine faint surveys whose data are compiled
here searched within a few degrees of the ecliptic
plane (Table~\ref{datatable}), where there is no bias against
finding Classical TNOs, with one exception:
\citet{fuentes10} searched fields $<$5~degrees
from the ecliptic. This
should technically not introduce a bias
against Classical objects, since our definition of
Classical is those objects having inclinations
$<$5~degrees. Furthermore, the \citet{fuentes10}
search
contributes only a single R$>$26 object to our
catalog, and thus can contribute little overall
bias to our result.

\section{Discussion}

\begin{figure}
\begin{center}
\includegraphics[angle=270,scale=0.5]{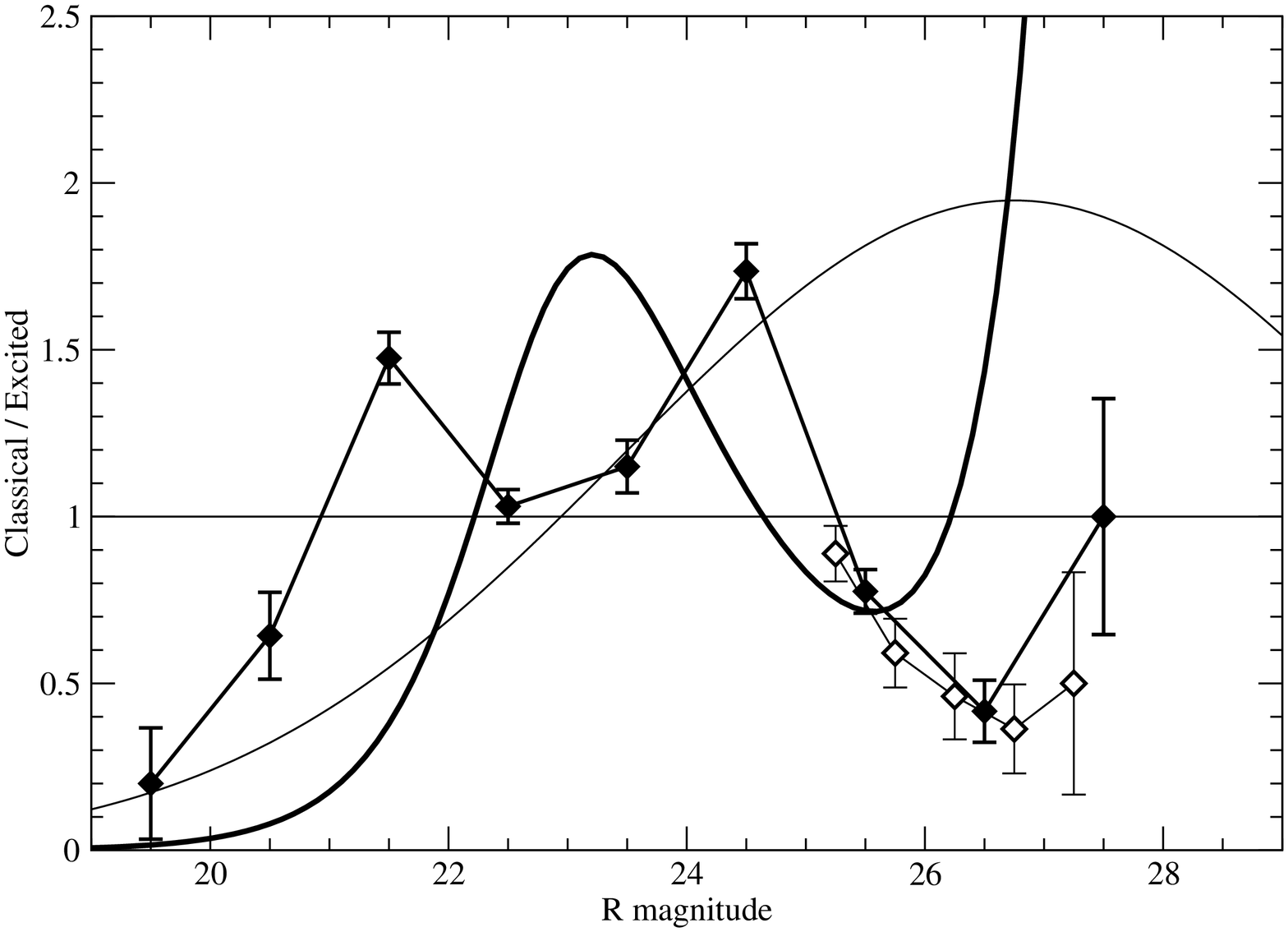}
\end{center}
\caption{Ratio of Classical TNOs
to Excited TNOs as a function of R magnitude.
Shown here is the predicted ratio of 
Classical to Excited TNOs from B04 (curved lines).
The B04 double-power law is shown by
the thick curved line, while the B04 rolling-index
power law is shown by the thin curved line.
The compiled data presented here is shown as diamonds
with bin sizes of 1~mag (dark diamonds)
and 0.5~mag (white diamonds, shown only
for R$>$25).
The error bars are calculated as $\sqrt{N}/N$.
Faintward of R=26, the data clearly do not
follow the B04 curves: rather than Classical
TNOs dominating, Excited TNOs instead dominate.
\label{ratio}}
\end{figure}

Our data compilation shows that Excited
objects dominate the faint TNO population.
Why, then, did B04 find only Classical objects
in their survey? At R=[26.3,27.5,27.8] (the 
magnitudes of the three B04-discovered
TNOs \citep{erratum}), the ratio of 
Classical to Excited TNOs is approximately
[0.4,0.5,$\sim$0.5] (Figure~\ref{ratio}) and the fraction of
all TNOs that are Classical
(that is, N(Classical)/(N(Classical)+N(Excited))) is therefore
approximately
[0.3,0.3,0.3].
The probability of finding these three 
Classical objects is therefore
0.3$^3$, or around 3\%.
To translate this binomial probability
to a conceptual gaussian equivalent, this
is about a 2.2$\sigma$ result, so not
drastically improbable.
This may be ample explanation.
The fact that B04 discovered only three
objects may indicate that small-number
statistics drove the 
B04 analysis.

The new data presented here shows that
the faint population is dominated by
Excited TNOs.
The much-cited ``Nice model'' describes a
Solar System-wide upheaval that occurred 
some 600~Myr after the formation of the
planets \citep{levison08}. During that event,
small bodies, including TNOs, were scattered
throughout the Solar System.
In particular, proto-TNOs that were found
within the planetary realm would have
been ejected to Excited orbits outside
the planets during this episode.
The
\citet{levison08} results broadly reproduce the
observed structure of the TNO region, including
the existence of both Excited and Classical objects.
However, the 
specific finding that Excited TNOs are
far more numerous at faint magnitudes
(which probably, but not certainly, correspond
to small sizes; see below)
is not explained
within the context of the Nice model.

Small TNOs are either the product of
planetesimal formation (creation) or
the result of collisional grinding
among larger TNOs (destruction).
Under the creation hypothesis, the ratio
between Classical and Excited TNOs records
their (possibly distinct) formation histories,
and the fact that Excited objects dominate
the bright and faint ends and not the middle
is difficult to explain unless the two
populations formed with very different
size-frequency distributions.
Alternately, small Excited TNOs may
be created by the collisional destruction of 
larger Excited TNOs, which predicts
that TNOs an order of magnitude
larger than the Excited-dominated
regime --- that is, TNOs around
R=23--24 --- should be dominated
by Classical objects. In other words,
the ``excess'' of small Excited TNOs at R$>$26
results from a collisional ``deficit''
of medium-sized Excited TNOs at R=23--24, which
matches the observational data.
Higher rates of collisional destruction
among Excited than Classical TNOs could
be explained simply by the fact that
Excited TNOs have larger mutual velocities
in the present day Kuiper Belt due to
their excited orbits; a similar result
has been found for the main asteroid
belt \citep{terai,obriendps}.
Alternately,
a higher collisional destruction rate
for Excited TNOs could be a result of
their hypothesized
formation closer to the Sun (according to
the Nice model), which would provide both
shorter evolutionary timescales
and higher mutual velocities. In either case,
the size distribution of Excited TNOs could
include subtle features that derive from 
collisional evolution.

Another possible explanation exists.
In general, Classical objects have distances 40--45~AU.
Excited objects have a wider range of heliocentric distance,
and dominate the populations both closer than and
farther than 40--45~AU. Therefore, one interpretation
of the data shown in Figure~\ref{ratio} is
that the ratio of Classical and Excited TNO
sizes is relatively constant with size,
and that the observation that the Excited population
dominates the bright and faint ends reflects simply
the change in apparent brightness due to the objects'
distances from the Earth. As a complicating factor,
Classical TNOs appear to have higher albedos than
other dynamical subgroups \citep{stans08,brucker}.
Thus, Classical TNOs of a given size appear fainter
than Excited TNOs of the same size.

There
are three very faint
Classical TNOs not shown in Figure~\ref{ratio}
(2003~BF91, R=27.5, i=1.49;
2003~BH91, R=27.8, i=1.97; and
c5a4, R=28.5, i=0) because
there are no Excited objects in these
sparsely-populated bins, giving a ratio
of infinity.
Thus, in our faintest bin, we are as data-limited
as B04 was --- interestingly, again with only
Classical objects in the faintest bin --- and no significant conclusion
can be drawn.
New data that pushes the faint limit
another 1~or 2~magnitudes would
allow us to constrain this ``very faint'' population
and help determine whether the differences in the
two populations are dynamical (collisional
destruction) or geometrical (similar size distributions
at different locations, modulated by
albedo effects).
Although very 
difficult to obtain, new data
at these very faint magnitudes will
be quite powerful in constraining the history
of the outer Solar System.


\acknowledgments

We acknowledge support from 
Hubble Space Telescope award
HST-AR-11778
and NASA Planetary Astronomy 
award NNX09AE76G. We thank
Mike Brown, JJ Kavelaars, and an anonymous
referee for useful comments.

\clearpage



\clearpage







\clearpage

\begin{deluxetable}{llllll}
\rotate
\tablecaption{Compiled survey data \label{datatable}}
\tablewidth{0pt}
\tablehead{
\colhead{Survey} & \colhead{Telescope} & \colhead{Sensitivity} &
\colhead{Area (deg$^2$)} & \colhead{$|$Ecl.\ lat.\ (deg)$|$} & \colhead{$N(R>26)$}}
\startdata
\citet{cb} & Keck & 27.9 (V) & 0.01 & 0.54 & 1 \\
\citet{gladman} & CFHT, VLT & 25.9 (R), 26.7 (R) & 0.27, 0.012 & 2.64,
1.04 & 3, 0 \\
\citet{abm} & NOAO 4-m & 24.9 -- 25.4 (R) & 2.3 & 0.02--1.93 & 0 \\
\citet{b04,erratum} & HST & 29.2 (F606W) & 0.02 & 1.48 & 3 \\
\citet{elliot05} & NOAO 4-m & 22.5 (VR) & 550 & $<$6 & 0 \\
\citet{fraser08} & CFHT, CTIO 4-m & 25.1 (VR) -- 26.4 (g$^\prime$) &
3.0 & 0.06--0.76 & 3 \\
\citet{fuentes08} & Subaru & 25.7 (R) & 2.8 & 0.13--1.88 & 0 \\
\citet{fraser09} & Subaru & 26.9 (r$^\prime$) & 0.26 & 0.29 & 18 \\
\citet{fuentes09} & Subaru & 26.8 (R) & 0.26 & 0.6 & 10 \\
\citet{fuentes10} & HST & 26.1 (R) & 0.45 & $<$5 & 1 \\
\enddata
\tablecomments{We list here the 10~surveys whose
data we compile in this work. The 50\% sensitivities
in the indicated filter are given along
with the searched area, absolute value of the ecliptic
latitude of the search, and number of objects R$>$26 that
were detected.
We used V-R=0.6 \citep{tr03};
VR-R=0.03 \citep{fraser08};
g$^\prime$-R=0.95 \citep{fraser08}; and
r$^\prime$-R=0.26 \citep{fraser08} to convert
to R magnitudes.
}
\end{deluxetable}





\end{document}